\documentclass{PoS}
\usepackage{nico}

\title{Weak Matrix Elements of Beyond the Standard Model $\Delta s=2$ four-quark operators 
from $n_f=2+1$ Domain-Wall fermions}

\ShortTitle{BSM $\Delta s=2$ four-quark operators from $n_f=2+1$ Domain-Wall fermions}

\author{\speaker{N.~Garron} \\
{School of Mathematics, Trinity College, Dublin 2, Ireland}\\
        E-mail: \email{ngarron@maths.tcd.ie}}
\author{{P.~A.~Boyle, R.~J.~Hudspith}\\
{SUPA, School of Physics, The University of Edinburgh, Edinburgh EH9 3JZ, 
United Kingdom}\\
 E-mail: \email{paboyle@ph.ed.ac.uk}\\
 E-mail: \email{R.Hudspith@sms.ed.ac.uk}
}
\author{{A.~T.~Lytle}\\
{School of Physics and Astronomy, University of Southampton, Southampton SO17 1BJ, United Kingdom\\}
 E-mail: \email{a.t.lytle@soton.ac.uk}
}

\author{{The RBC and UKQCD Collaborations}\\  \flushright{TCDMATH 12-13 \,Edinburgh 2012/24}}

\abstract{
We report on our computation of the hadronic matrix elements of the four-quark operators needed 
for the study of $K^0-{\bar K^0}$ mixing beyond the Standard Model (SM). 
We consider $n_f=2+1$ Domain-Wall fermions on Iwasaki gauge action 
with lightest unitary pion of $290\, \MeV$ and 
a single lattice spacing $a\sim0.086\, \fm $.
The renormalization is performed non-perturbatively through the RI-MOM scheme 
and our results are converted perturbatively to $\msbar$. 
We have estimated the various systematic errors. Our results confirm a 
previous quenched study, where large ratios of non-SM to SM matrix elements 
were obtained.}

\FullConference{The 30 International Symposium on Lattice Field Theory - Lattice 2012,\\
		June 24-29, 2012\\
		Cairns, Australia}

\begin{document}

\section{ Introduction}
Direct and indirect CP violation in kaon systems  
are being currently investigated 
by the RBC-UKQCD with a framework of $n_f=2+1$ dynamical flavours of Domain-Wall.  
Exciting results concerning direct CP violation obtained 
though simulations of the decay of a kaon into two pions
have been reported in~\cite{Blum:2011ng,Blum:2012uk,Blum:2011pu,Boyle:2012ys}.
Important progress has also been achieved concerning indirect CP violation,
and in particular our computation of the non-SM contributions 
to neutral kaon mixing has been recently published in~\cite{Boyle:2012qb}.
We summarise here our strategy and the current status of our analysis.
We also mention that other collaborations have recently reported on 
similar studies at this conference~\cite{
Bailey:2012dz, Bailey:2012bh, Carrasco:2012dd}

After performing an operator-product-expansion, neutral kaon mixing can described 
by a generic $\Delta s=2$ Hamiltonian of the form
\be
\label{eq:H}
H^{\Delta s=2} = \sum_{i=1}^{5} \, C_i(\mu) \, O_i^{\Delta s=2}(\mu) \,,
\ee
where $\mu$ is a renormalization scale.
The Wilson coefficients $C_i$, which encode the short-distance effects, 
depend on the new physics model under consideration. The long-distance effects
are factorised into the matrix elements of the four-quark operators $O_i^{\Delta s=2}$
given here in the so-called SUSY basis\footnote{We discard the parity odd operators
since they are not relevant in the present case.}
~\cite{Gabbiani:1996hi}
\bea
\label{eq:OBSM}
O_1^{\Delta s=2} &=&
(\overline s_\alpha \gamma_\mu (1-\gamma_5) d_\alpha)\,
(\overline s_\beta  \gamma_\mu (1-\gamma_5) d_\beta) \, \nonumber,\\
O_2^{\Delta s=2}&=&
(\overline s_\alpha (1-\gamma_5) d_\alpha)\,
(\overline s_\beta  (1-\gamma_5) d_\beta)\nonumber,\\
\label{eqO3}
O_3^{\Delta s=2}&=&
(\overline s_\alpha  (1-\gamma_5) d_\beta)\,
(\overline s_\beta   (1-\gamma_5) d_\alpha)\nonumber,\\
O_4^{\Delta s=2}&=&
(\overline s_\alpha  (1-\gamma_5) d_\alpha)\,
(\overline s_\beta   (1+\gamma_5) d_\beta)\nonumber,\\
\label{eqO5}
O_5^{\Delta s=2}&=&
(\overline s_\alpha  (1-\gamma_5) d_\beta)\,
(\overline s_\beta   (1+\gamma_5) d_\alpha).
\eea
In the SM case ($i=1$) it is conventional to introduce the kaon bag
parameter $B_K$, which measures the deviation of the SM matrix element from its
vacuum saturation approximation (VSA)
\be
B_K = {  \langle \Kb| O_1^{\Delta s =2} | \K \rangle  \over { {8\over3} m_K^2 f_K^2 }} \;
\;.
\ee
Where the normalisation for the decay constant is such that
$f_{K^-}\;=\:156.1 \MeV$. 
Several normalisations for the BSM operators ($i>1$) have been proposed in the literature,
see for example~\cite{Donini:1999nn}, 
in this work we follow~\cite{Babich:2006bh} 
and define the ratios
\be
\label{eq:R}
R_i^{\rm BSM}  = 
\left[ {f_K^2 \over m_K^2} \right]_{\rm expt}
\left[ {m_P^2 \over f_P^2} { \langle \Pb| O_i^{\Delta s =2} | \P \rangle \over \langle \Pb|
O_1 | \P \rangle }\right]_{\rm latt} \;,
\ee
where $P$ is a pseudo-scalar particle of mass $m_P$ and decay constant $f_P$.
The term $\left[\:\right]_{\rm latt}$ is obtained from our lattice simulations
for different values of  $m_P$.
For completeness we will also give the BSM bag parametrisation, defined as
(where $N_{2,\ldots,5}={\frac 53,-\frac 13, -2, -\frac 23} $)
\cite{Conti:1998ys},
\begin{equation}\label{eq::BSM_BAG}
B_i \;=\; -\frac{\langle \Kb | O_i^{\Delta s =2} | \K \rangle}{N_i \langle \Kb |
\bar{s}\gamma_5 d |0\rangle \langle 0 | \bar{s}\gamma_5 d| \K
\rangle},
\quad i=2,\ldots,5
.
\end{equation}

\section*{Computation details}\label{sec:strategy}
This computation is performed on 
$32^3\times64\times16$ 
Iwasaki gauge configurations with an inverse lattice spacing 
$a^{-1}=2.28(3)\,\GeV$, corresponding to $a\sim0.086\,\fm$
\footnote{Note that this value was recently updated to 
$a^{-1}=2.31(4)\, \GeV$ \cite{:2012yc}.}. 
We have three different values of the 
light sea quark mass $am_{\rm light}^{\rm sea}=0.004, 0.006, 0.008$
corresponding to unitary pion masses
of approximately $290, 340$ and $390 \;\MeV$ respectively. 
The simulated strange sea quark mass is $am_{\rm strange}^{\rm sea}=0.03$,
which is close to its physical value $0.0273(7)$. 
For the main results of this work, we consider only unitary light quarks, 
$am_{\rm light}^{\rm valence}=am_{\rm light}^{\rm sea}$, 
whereas for the physical strange we interpolate between the unitary 
($am_{\rm strange}^{\rm valence}= am_{strange}^{\rm sea}=0.03$) and the partially
quenched $(am_{\rm strange}^{\rm valence}=0.025) $ data. 

\begin{figure}[!h]
\begin{center}
\begin{tabular}{cc}
\includegraphics[width=7cm]{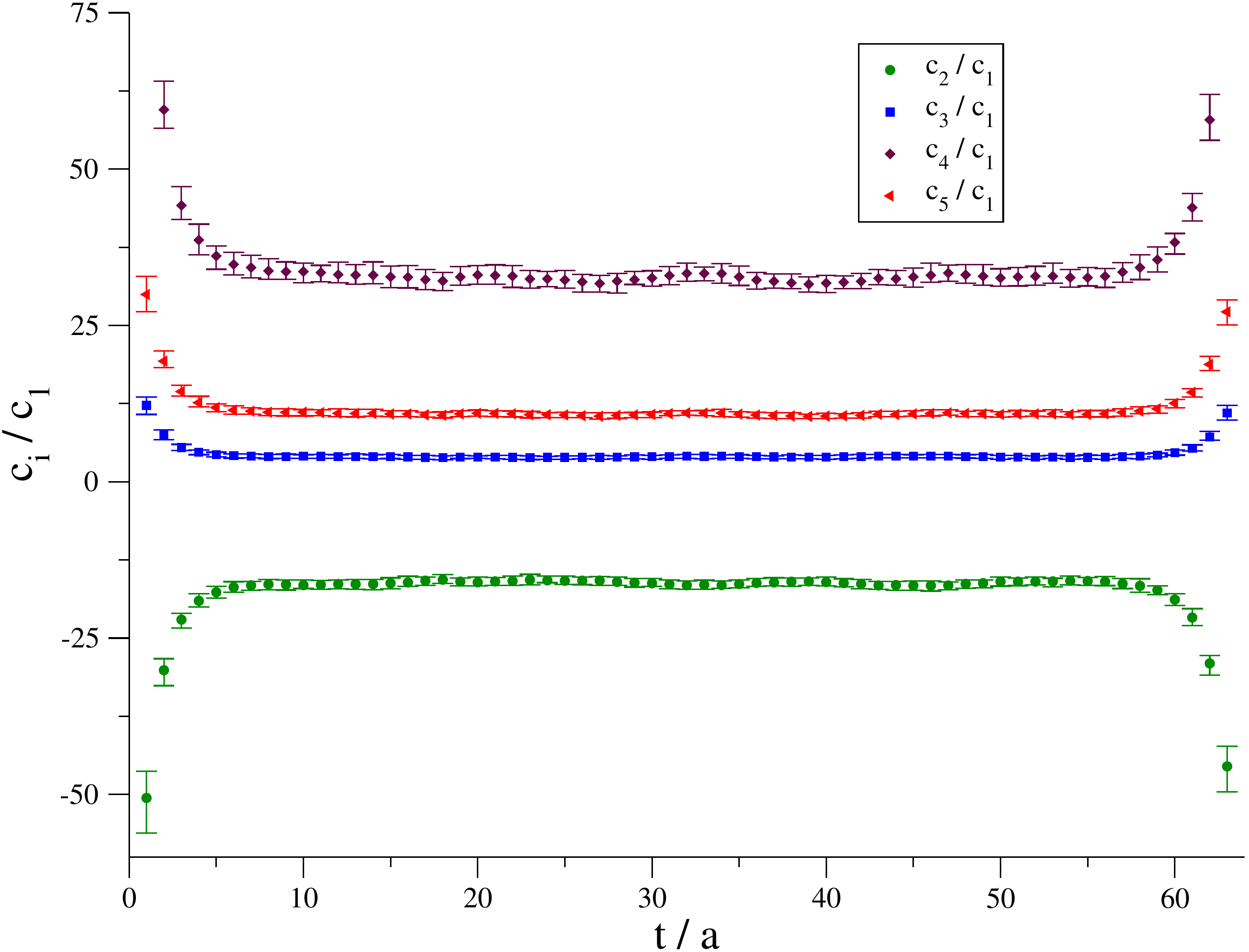} &
\includegraphics[width=7cm]{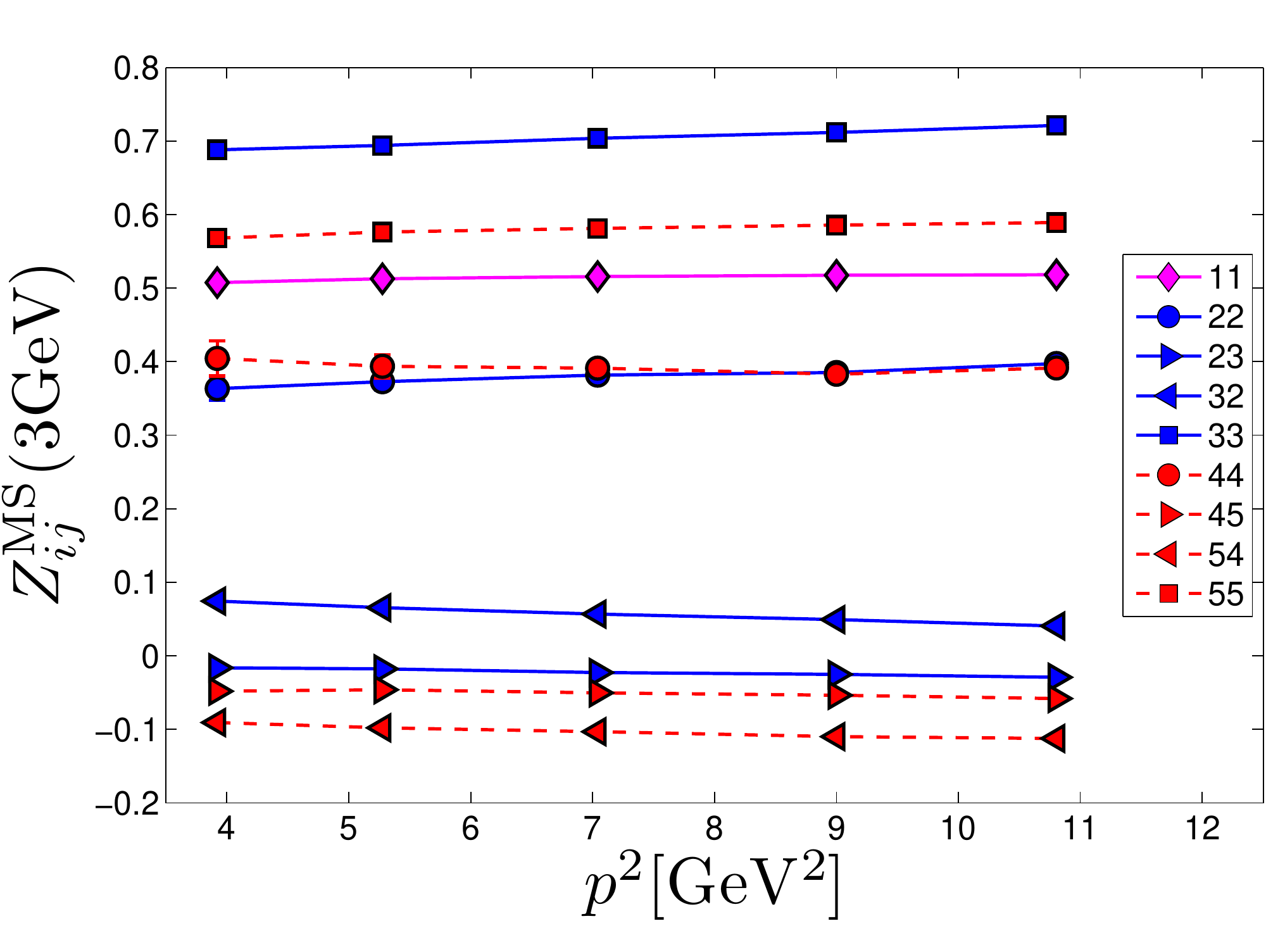}
\end{tabular}
\caption[]{{\em Left:} ratios of the bare three point functions from
which we extract $R_i^{\rm BSM}$.
Results are shown for our lightest simulated unitary kaon.\\
{\em Right:} renormalization factors of the four-quark operators.
At each value of the simulated momentum $p$, we run to the scale of $3\GeV$ and 
convert to $\msbar$. The remaining scale dependence 
can be imputed to the truncation of the perturbative expansion.
We show only the $Z$-factors allowed by chiral symmetry.
\vspace{-0.5cm}}
\label{fig:plateaux}
\end{center}
\end{figure}

The procedure for the evaluation of the two-point functions and of the three point function 
is fairly standard (in particular, we have used Coulomb gauge fixed wall sources to obtain very good
statistical precision).
We define the
three point functions
$c_i = \la \Pb(t_f) O^{\Delta S=2}_i(t) \Pb(t_i) \ra$ 
and from the asymptotic Euclidean time behaviour 
of the ratios of three point-functions $c_i/c_1$
(we fit these ratios to a constant in the interval $t/a=[12,52]$ \footnote{From
figure~\ref{fig:plateaux} we deduce that we have reached the asymptotic region 
in this range for each operator insertion.}) we 
obtain the bare matrix elements of the four-quark operators 
normalised by the SM one:
$\left[ \langle \Pb| O_i^{\Delta S=2} | \P \rangle / \langle \Pb| O_1^{\Delta S=2} | \P \rangle\right]^{\rm bare}$ .
In figure~\ref{fig:plateaux} (left), we show the corresponding plateaux 
obtained for our lightest unitary kaon 
$am_{\rm light}^{\rm sea} =  am_{\rm light}^{\rm valence} =0.004$,
$am_{\rm strange}^{\rm sea} = am_{\rm strange}^{\rm valence}=0.03$.
\\
\section*{Renormalization}\label{sec:renormalization}
The four-quark operators given in eq.~(\ref{eq:OBSM}) mix under
renormalization. With Domain-Wall fermions, the renormalization pattern
is the same as in the continuum (up to numerically irrelevant discretisation
effects). 
The SM operator 
$O_1^{\Delta s=2}$ belongs to a $(27,1)$ irreducible representation of 
$SU(3)_L \times SU(3)_R$
and renormalizes multiplicatively.
The BSM operators fall into two categories: 
$O_2^{\Delta s=2}$ and $O_3^{\Delta s=2}$ transform like 
$(6,\overline 6)$ 
and mix together.
Likewise $O_4^{\Delta s=2}$ and $O_5^{\Delta s=2}$ belong to $(8,8)$
and also mix. 

We perform the renormalization of the four-quark operators $O_i^{\Delta s=2}$
non-perturbatively in the RI-MOM scheme~\cite{Martinelli:1994ty}.
By using both momentum sources~\cite{Gockeler:1998ye} and partially
twisted boundary conditions, we obtain 
smooth functions of the external momentum with 
very good statistical accuracy~\footnote{More details about the computation of the
renormalization factors can 
be found in~\cite{Boyle:2011kn,Boyle:2011cc,Arthur:2011cn}}. 
Although we have also performed this computation in 
a non-exceptional intermediate scheme,
we quote here the results obtained via the RI-MOM scheme
because only in this case are the conversion factors to $\msbar$
(computed in continuum perturbation theory) available for the whole set of operators.
We choose to impose the renormalization conditions at 
$\mu=3\,\GeV$, the conversion between $3$ and $2\,\GeV$
can be found in the appendix.

We observe that the effects of chiral symmetry breaking are not 
completely negligible, even at $3\,\GeV$~\cite{Boyle:2011kn}.
Therefore we must assess a systematic error to the mixing of operators
of different chirality (see next section).
We have checked that in a non-exceptional scheme this small 
chirally forbidden mixing is strongly reduced and becomes numerically irrelevant 
at $3\,\GeV$~\cite{Boyle:2011kn,Wennekers:2008sg}. Thus we conclude 
that this effect is a physical manifestation of the infrared behaviour of the exceptional 
intermediate scheme. 
The results for the chirally allowed renormalization factors $Z_{ij}^{\msbar}(3\GeV)$
are shown in figure~\ref{fig:plateaux} (right).
They relate the bare four-quark operators to the 
renormalized ones through the usual relation 
($Z_q$ is the renormalization factor of the quark wave 
function and cancels in the ratios)
\be
O_i^{\Delta s=2,\rm \msbar}({3\GeV}) = {Z_{ij}\over Z_q^2}^{\msbar}\!\!\!\!({3\GeV})\,
{O_j^{\Delta s=2,\rm bare}} \;.
\ee
\section*{Physical results and error estimation}\label{sec::Results}
\begin{figure}[t]
\begin{center}
\includegraphics[width=9cm]{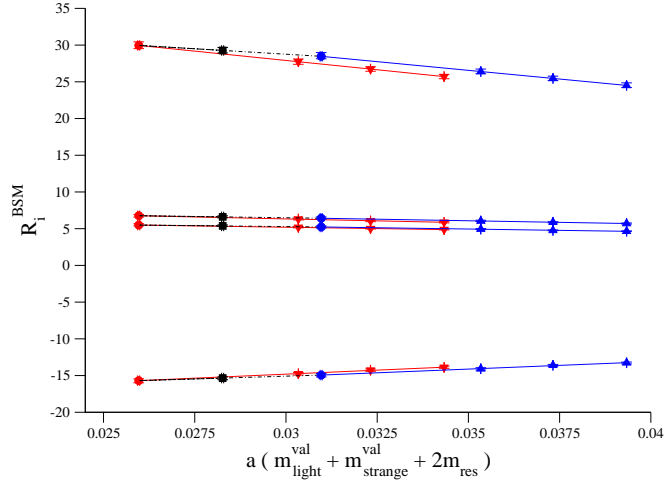}
\caption[]{Renormalized BSM ratios $R^{\rm BSM}_i$ in function of the bare 
valence quark mass.
We show the three unitary light quarks for both 
the unitary strange (upward blue triangles) and the partially
quenched strange (downward blue triangles), 
together with their extrapolations 
in the light sector (blue and red circles) and their interpolation 
to the physical kaon mass (black squares).
\vspace{-0.5cm}
}
\label{fig:extrap}
\end{center}
\end{figure}

Once the bare ratios have been renormalized, we extrapolate them to the 
physical kaon mass. 
The chiral functional form of the BSM operators are discussed for example in 
~\cite{Becirevic:2004qd, Detmold:2006gh, Bailey:2012dz}.
Since we find that the $R^{\rm BSM}$'s exhibit a very mild quark mass dependence
(see figure~\ref{fig:extrap}),
we take the results obtained by a simple Taylor expansion  
as our central values.

Our final results 
are the $R_i^{\rm BSM}$ quoted in $\msbar$ at $3\, \GeV$ 
given in table \ref{tab:results}. 
For completeness, we also convert these to the 
BSM bag parameters, using eq.~(\ref{eq::BSM_BAG}).
We also note that, using the same framework, the SM contribution is found to 
be 
$B_1 = B_K = 0.517 (4)_{\rm stat}$
in the $\msbar$ scheme at $3\,\GeV$, 
whereas a continuum value of 
$0.529(5)_{\rm stat} (19)_{\rm syst}$ was quoted in~\cite{Aoki:2010pe}. 
The difference comes from the fact that a different intermediate 
scheme was used in~\cite{Aoki:2010pe} (such a difference is accounted for in
our estimation of the systematic errors). 
From the same reference, the discretisation 
effects for $B_K$ on this lattice are seen to be of the order of $1.5\%$.
Since we have only one lattice spacing for the BSM ratios,
we make the assumption that the discretisation errors are 
of the same size as those affecting $B_K$, and estimate a $1.5\%$ error
to all the different operators.
This might appear like a crude estimate, but this effect is expected to be 
sub-dominant compared to other sources of systematic errors.
The next systematic error (called ``extr.'') represents the spread 
of the results obtained from different extrapolation strategies
to the physical point.
The systematic error associated with the non-perturbative renormalization
(NPR) has been estimated from the breaking of chiral symmetry.
The mixing of the $(6,\bar 6)$ with the $(8,8)$ operators 
is forbidden by chiral symmetry, but is likely to be enhanced 
by the exchange of light pseudo-scalar particles.
As the matrix element of $O_4^{\rm \Delta s=2}$
is numerically large, this non-physical mixing has an
effect on $O_2^{\rm \Delta s=2}$ and $O_3^{\rm \Delta s=2}$ 
of the order of $8-9\%$.
This unwanted infrared effect is absent if a non-exceptional 
scheme is used.
The last error we quote (``PT'') arises from the matching 
between the intermediate RI-MOM scheme and $\msbar$,
which is performed at one-loop order in perturbation 
theory~\cite{Ciuchini:1993vr,Buras:2000if}
in the three-flavour theory.
The associated error is obtained by taking half the difference between the leading
order and the next to leading order result~\footnote{
To obtain $\alpha_s$ at $3\,\GeV$ in the three-flavour theory,
we start from $\alpha_s(M_Z) =0.1184$ 
\cite{Nakamura:2010zzi}, we use the four-loop 
running~\cite{vanRitbergen:1997va,Chetyrkin:1997sg} 
to compute the scale evolution down 
to the charm mass, while changing the number of flavours when crossing a threshold,
and then run up to $3\,\GeV$ in the three-flavour theory.}.
We note that this error is one of the dominant ones in our budget,
and we expect this error to be reduced by an important factor 
if a non-exceptional scheme were used, since the latter are known to converge faster
in perturbation theory. 
We neglect the finite volume effects which have been found to be small
in~\cite{Aoki:2010pe}, as one can expect from the value of $m_\pi L\sim 4$ 
for our lightest pion mass $m_\pi\sim 290\, \MeV$.
\begin{table}[t]
\begin{center}
\begin{tabular}{ c | c  c  ||  c  c  c  c  c | c }
\hline
i & $R_i^{\rm{BSM}}$ & $B_i$ 
& stat. & discr. & extr. & NPR & PT  & total
\\
\hline
$2$  & $-15.3 (1.7)$  & 0.43 (5)  & $1.3$  & $1.5$ & $4.0$  & $9.4$ & $4.7$   & $11.3$ \\
$3$  & $  5.4 (0.6)$  & 0.75 (9)  & $2.0$  & $1.5$ & $3.9$  & $7.8$ & $7.6$   & $12.0$  \\ 
$4$  & $ 29.3 (2.9)$  & 0.69 (7)  & $1.3$  & $1.5$ & $4.1$  & $3.0$ & $8.2$   & $9.8$  \\ 
$5$  & $  6.6 (0.9)$  & 0.47 (6)  & $2.1$  & $1.5$ & $3.8$  & $3.2$ & $12.6$  & $13.8$
\end{tabular}
\caption{Final results of this work: the first two columns 
show the ratios $R_i^{\rm BSM}$ and the corresponding bag
parameters $B_i$ in $\msbar$  at $3\,\GeV$,
together with their total error, combining systematics and statistics.  
In the remaining columns, we give our error budget 
for the $R^{\rm BSM}$, 
detailing the contributions in percentage of the different sources 
of systematics (see text for more details).
\vspace{-0.5cm}
}
\label{tab:results}
\end{center}
\end{table}

%
\section*{Conclusions}\label{sec::Conclusions}
We have computed the electroweak matrix elements of the $\Delta s=2$ 
four-quark operators which contribute 
to neutral kaon mixing beyond the SM. 
Our work improves on other studies by using 
$n_f=2+1$ flavours of dynamical chiral fermions.
We confirm the effect seen in a previous quenched 
computation \cite{Babich:2006bh}, 
where large enhancements of the non-standard matrix elements were observed. 
The errors quoted in this work are of the order of $10\%$. 
We note that the main limitation of this study comes from the lack of matching factors between 
non-exceptional renormalization schemes (such as SMOM) and $\msbar$. 
Once these factors are
available, we expect to reach a precision better than $5\%$.
We have started to repeat this computation with another lattice spacing 
in order to have a better handle on the discretisation effects.

\section*{Appendix}
We have computed the non-perturbative scale evolution of the $R^{\rm BSM}$'s
between $3$ and $2\,\rm GeV$, and then converted the results to $\msbar$ 
using one-loop perturbation theory~\cite{Ciuchini:1993vr,Buras:2000if}:
\bea
&&
U^{\msbar}(2\,{\rm GeV}, 3\,{\rm GeV})
= \qquad \qquad\qquad \qquad  \nn\\
&& 
\left(
\begin{array}{r r r r r }
    1       &     0 &      0 &        0 &        0 \\
         0  &  0.87 &   0.02 &        0 &        0 \\
         0  &  0.09 &   1.09 &        0 &        0 \\
         0  &       0 &        0 &   0.86 &  -0.01 \\
         0  &       0 &        0 &  -0.03 &   0.98
\end{array}
\right) \;.
\eea
Our conventions are such that 
\be\
R^{\rm BSM}({2\,\GeV}) = U^{\msbar}(2\,{\rm GeV}, 3\,{\rm GeV}) R^{\rm BSM}(3\,\GeV) \,.
\nn
\\
\ee\
\section*{Acknowledgements}
We would like to thank the organisers of Lattice 2012 for such an enjoyable conference.
We are grateful to our colleagues of the RBC and UKQCD collaborations.
We acknowledge F.~Mescia and S.~Sharpe for discussions
and the members of the ETM Collaboration for discussing their
results before publication. 
R.J.H. acknowledges the STFC grant ST/G000522/1 and 
the EU grant 238353 (STRONGnet).
A.T.L akcnowledges the STFC grant ST/J000396/1. 
The University of Southampton's Iridis cluster is funded by STFC grant ST/H008888/1.
Renormalization was performed using STFC funded DiRAC resources.

\bibliography{biblio}{}
\bibliographystyle{h-elsevier}

\end{document}